\documentclass[aps,reprint,amsmath,amssymb,superscriptaddress,floatfix,twocolumn
,nofootinbib 
]{revtex4-1}
\usepackage{natbib}
\usepackage{graphicx,dcolumn,lipsum,placeins}
\usepackage[utf8]{inputenc}
\usepackage[T1]{fontenc}
\usepackage[all]{nowidow}
\usepackage{physics}
\usepackage{hyperref}
\usepackage[bottom]{footmisc}
\usepackage{booktabs}
\usepackage[en-IE]{datetime2}
\DTMlangsetup[en-IE]{ord=raise}
\usepackage[dvipsnames,table,xcdraw]{xcolor}
\graphicspath{{images/}{../images/}{images/1}{images/2}}
\usepackage[linesnumbered,ruled,vlined]{algorithm2e}
\DeclareUnicodeCharacter{25A0}{
	\rule[1pt]{5.25pt}{3pt}}
\usepackage{hyperref}[backref]
\hypersetup{
	colorlinks = true,
	allcolors = {blue},
}

\usepackage{orcidlink}
\begin{document}
\def\r{.95}
\title{Crosstalk-Induced Row Hammer Attacks on Quantum Random Access Codes}
\author{Fernando Almaguer-Angeles$^\star$\orcidlink{0000-0002-3521-7256}}
\email{fernando.almaguerangeles@phdstud.ug.edu.pl}
\affiliation{International Centre for Theory of Quantum Technologies,
University of Gdańsk, Profesor Marii Janion 7, 80-309 Gdańsk, Poland}

\author{Victor F. dos Santos$^\star$\orcidlink{0009-0009-0319-4852}}
\email{victorfds997@gmail.com}
\affiliation{Physics Department,
Federal University of Santa Maria, 97105-900,
Santa Maria, RS, Brazil}

\author{Jonas Maziero\orcidlink{0000-0002-2872-986X}}
\email{jonas.maziero@ufsm.br}
\affiliation{Physics Department,
Federal University of Santa Maria, 97105-900,
Santa Maria, RS, Brazil}

\author{Marcin Pawłowski\orcidlink{0000-0002-8611-947X}}
\email{marcin.pawlowski@ug.edu.pl}
\affiliation{International Centre for Theory of Quantum Technologies,
University of Gdańsk, Profesor Marii Janion 7, 80-309 Gdańsk, Poland}

\author{Akshata Shenoy H.\orcidlink{0000-0002-6703-8383}}
\email{akshata.shenoy@ug.edu.pl}
\affiliation{International Centre for Theory of Quantum Technologies,
University of Gdańsk, Profesor Marii Janion 7, 80-309 Gdańsk, Poland}
\date{\today}

\begin{abstract}
Quantum processors remain susceptible to crosstalk-induced effects despite the continuous evolution of their underlying hardware architectures. In this work, we experimentally investigate the effectiveness of crosstalk-induced attacks across six IBM superconducting quantum processors and one IQM processor. Our results demonstrate that such attacks can be launched across all the platforms, due to persistance of crosstalk, although their effectiveness varies significantly between devices. The probability of a measurement outcome flip in the victim qubits reaches a maximum value of $60\%$. Also, it is seen that processors sharing identical coupling graphs exhibit different vulnerability profiles, indicating that connectivity alone is insufficient to predict attack efficiency. Finally, to demonstrate the practical relevance of these attacks, we study their impact on quantum random access codes (QRACs), observing a degradation in their average success probabilities.

\end{abstract}

\maketitle
\begingroup
\renewcommand{\thefootnote}{$\star$}
\footnotetext{These authors contributed equally to this work.}
\endgroup

\section{Introduction}
\label{sec:introduction}

Quantum processors have advanced considerably in the recent years, with experimental demonstrations improving their computational power and enabling increasingly complex quantum computations \cite{Preskill2018,Kim2023}. Nevertheless, reliable quantum computation requires not only accurate individual gate operations but also control over unintended interactions between qubits. Ideally, quantum operations performed on a particular qubit should not influence other neighboring qubits that are not intended to be affected by the operation. However, due to the hardware design and realization, unwanted interactions can cause these operations to inadvertently influence neighboring qubits, a phenomenon termed as crosstalk. Crosstalk remains a major source of errors in modern quantum processors, compromising the fidelity of quantum operations and processing tasks  \cite{Sarovar2020,Rudinger2021,Zhao2022,Kosen2024,heng2024estimating}. Experimental studies have investigated its characterization, physical mechanisms, and mitigation, including errors associated with simultaneous gate execution and coherent interactions between qubits \cite{Seif2024,Gross2024,perrin2024mitigating,Kang2025TimeAdaptiveCrosstalk,Wesdorp2026SimultaneousSingleQubitCrosstalk,ChenWang2026XYCrosstalk}.

It is interesting to note that crosstalk can potentially be exploited to implement a quantum analogue of the classical Row Hammer attack. Repeated application of quantum operations, such as quantum gates, on a qubit may induce unintended interactions with neighboring qubits through crosstalk, introducing errors and changing their states. As such, a quantum row hammer attack is launched by intentionally introducing disturbances in nearby victim qubits through repeated quantum operations via inter-qubit crosstalk in quantum processors. The applicability of the attack through cloud-based services, where hardware is shared between different users running their computational tasks, is known  \cite{Harper2025,ShubhaFarheen2025PulseLevelCrosstalk,Arellano2025QubitVise,MehraKalev2026DefendingCrosstalk}. These studies highlight that crosstalk is a potential vulnerability in quantum information processing \cite{AdhikariHalunen2025HardwareVulnerabilities} and motivate further investigation of its persistence and dependence on hardware realization.

Recent works have investigated crosstalk-induced effects in superconducting quantum processors using deliberately structured gate sequences~\cite{Almaguer2026, jakub24}. Furthermore, it is important to understand to what extent do such effects persist across different quantum processors and hardware architectures. By performing a systematic experimental study on different contemporary processors, characterization of the hardware to estimate the effects of such vulnerabilities is crucial. 
In particular, whether quantum processors with similar underlying connectivity exhibit comparable vulnerability profiles is of practical relevance for crosstalk-induced quantum attacks.

In this article, we investigate crosstalk-induced quantum row hammer effects in six IBM superconducting quantum processors, namely, Fez, Kingston, Marrakesh, Pittsburgh, Boston, and Miami, and the IQM Garnet processor. By employing two-qubit attack configurations, and adapting them to satisfy the connectivity constraints of the available quantum devices, the hardware characterization is performed. To characterize these effects at the hardware level, we measure the probability of observing the outcome $|1\rangle$ on a victim qubit initially prepared in the state $|0\rangle$ following the execution of each attack strategy. The measurement outcomes are compared against those obtained from a corresponding control experiment without launching any attack. This allows us to quantify the increase in the probability of an unintended transition from state $|0\rangle$ to $|1\rangle$. Our results show that crosstalk-induced measurement biases persist across all investigated platforms, although with varying magnitudes. In particular, processors sharing the same coupling graph can exhibit significantly different vulnerability profiles, demonstrating that connectivity alone is insufficient to predict the effectiveness of the attack.

Furthermore, we investigate the action of such attacks on $2\rightarrow1$ and $3\rightarrow1$ QRACs, where a qubit is prepared to encode two or three classical bits, respectively \cite{AmbainisLeung2009QRAC,Tavakoli2015,Wang2019EARAC}. A degradation in the optimal average success probability \cite{karthik25}, associated with the retrieval of the chosen bit by performing suitable measurements, is seen, as opposed to their ideal values of $0.854$ and $0.789$, respectively. Hence, the attack is characterized through both quantum-state reconstruction and average success probabilities. In addition, we compare two attack configurations, \textit{i.e.}, before and after state preparation, prior to the performance of a decoding measurement. This allows us to infer how the timing of the disturbance influences the encoded-state geometry and the retrieval process.

The remainder of this paper is organized as follows. Sec.~\ref{sec:exp} describes the experimental methods, including the quantum processors, attack configurations, and procedures used for both the hardware-level characterization and the QRAC experiments. Sec.~\ref{sec:results}  presents the hardware-level results for the IBM and IQM processors, examining the persistence and hardware dependence of crosstalk-induced attacks. Sec.~\ref{sec:qrac_attacks} investigates the impact of these attacks on the $2\rightarrow1$ and $3\rightarrow1$ QRACs, presenting the reconstructed quantum states and average success probabilities for the two attack positions. Finally, Sec.~\ref{sec:conclusion} summarizes the work and discusses their implications for the assessment of crosstalk-induced vulnerabilities in contemporary quantum processors.

\section{Experimental method}
\label{sec:exp}
\begin{figure}[t]
  \centering
  \includegraphics[width=\r\linewidth]{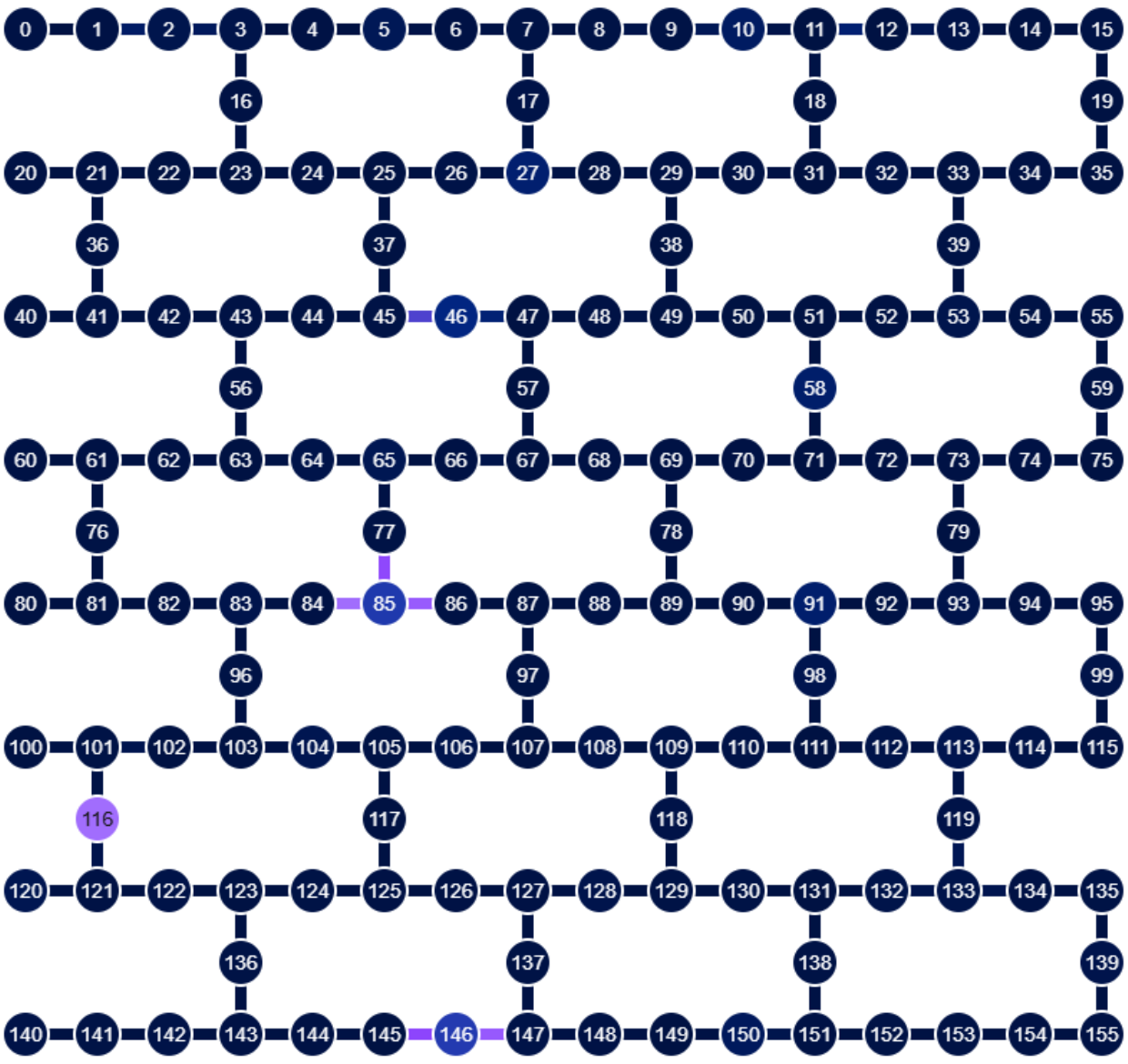}
  \caption{
Connectivity graph of the IBM quantum processors Fez, Kingston, Marrakesh, Pittsburgh, and Boston. Victim qubits 16, 37, 57, 78, 98, 119, 137, and 138 were attacked via a 30-CNOT gate configuration, as they possess six neighboring working qubits.
}
  \label{fig:IBM_top}
\end{figure}

\begin{figure}[ht]
  \centering
  \includegraphics[width=\r\linewidth]{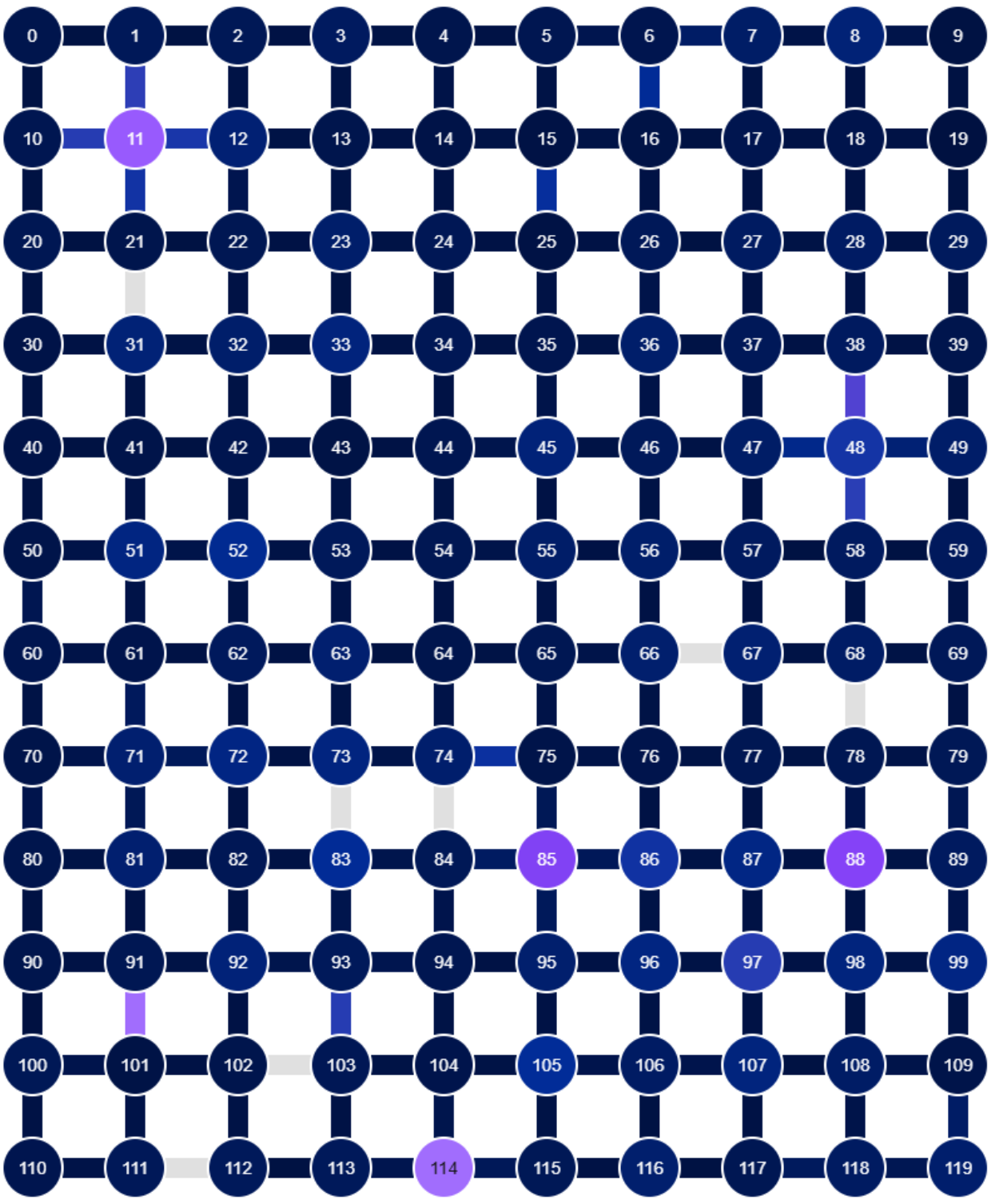}
  \caption{
Connectivity graph of the IBM Miami processor. Victim qubit 64 was selected for the experiments, as it allows the construction of complete square-like neighborhoods around it. Attack configurations consisting of one, two, three, and four concentric rings were investigated by applying all admissible bidirectional CNOT operations among qubits belonging to the selected neighborhood.
}
  \label{fig:miami}
\end{figure}

\begin{figure}[ht]
  \centering
  \includegraphics[width=\r\linewidth]{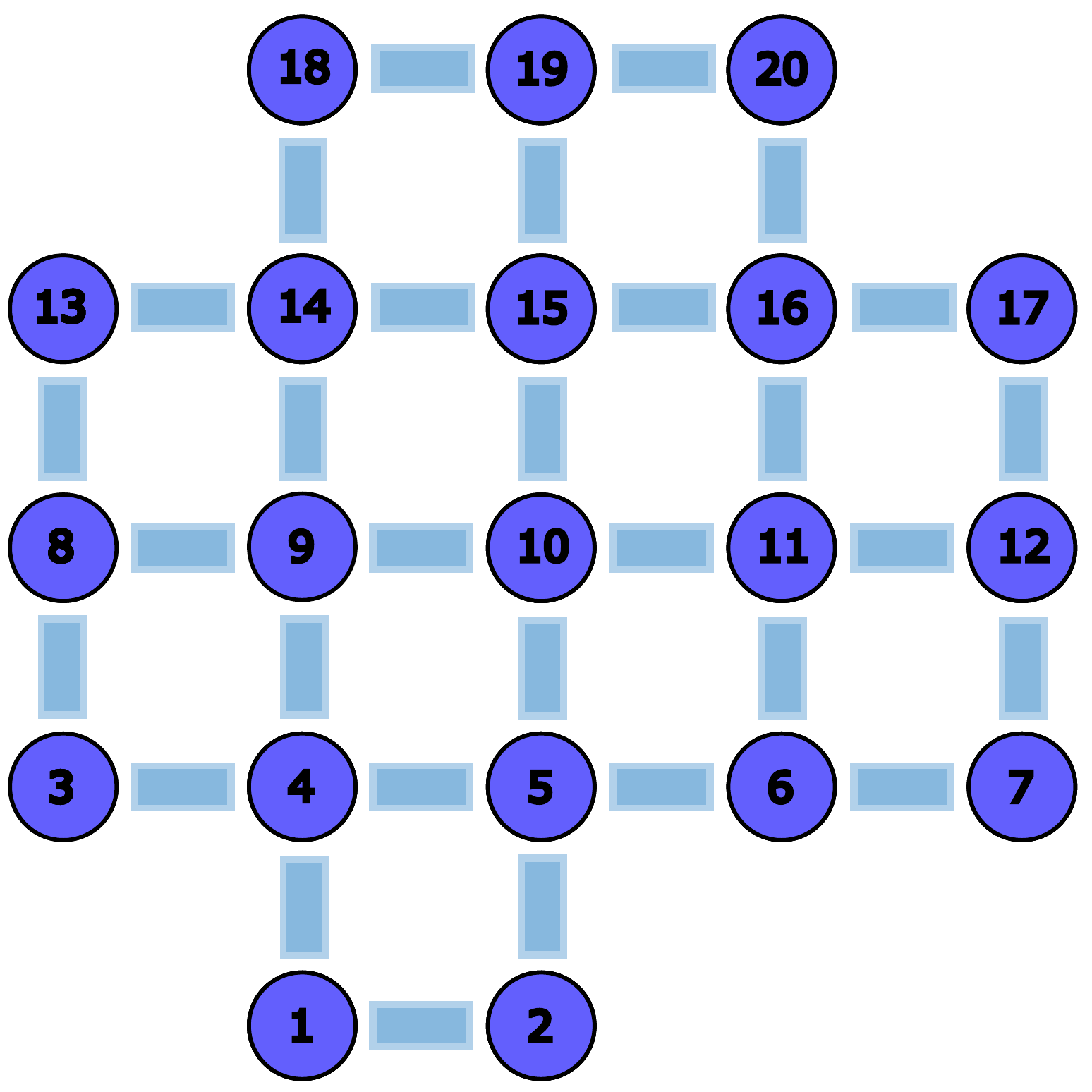}
  \caption{
Connectivity graph of the IQM Garnet processor. Victim qubits 9, 10, 11, and 15 were selected for the experiments, as they are the only qubits possessing a complete square-like neighborhood composed of eight surrounding qubits.
}
  \label{fig:iqm}
\end{figure}

Experiments were executed on superconducting quantum processors provided by IBM and IQM~\cite{KettererWellens2023}. Each circuit was executed with 40,000 repetitions on IBM devices, while 8,000 repetitions was performed on the IQM device. The IBM processors investigated in this work were Fez, Kingston, Marrakesh, Pittsburgh, Boston and Miami, corresponding to five 156-qubit processors sharing an identical heavy-hex connectivity graph~\cite{Chamberland2020HeavyHex,HetenyiWootton2024}
and one 120-qubit processor with a rectangular coupling layout. The only IQM processor available was Garnet, a 20-qubit superconducting device. For further details on qubit connectivity, refer to Figs.~\ref{fig:IBM_top}--\ref{fig:iqm}.

The hardware-level experiments were performed in two modes. In the control circuit mode, the victim qubit was prepared in the state $|0\rangle$ and measured in the computational basis, with the neighboring qubits being idle. The absence of an attack in this mode serves as a session-specific baseline for the victim qubit, accounting for inherent hardware imperfections and read-out errors ~\cite{Nation2021Scalable}. In the attack circuit ~\cite{AshSaki2020Crosstalk} mode, sequences of CNOT gates are applied to the neighboring qubits, leaving the victim outside this interaction, to induce crosstalk on the victim qubit~\cite{Gambetta2012Addressability}. The effectiveness of a given attack configuration is quantified by the probability of obtaining an outcome $1$ upon measuring the victim qubit. In the absence of crosstalk, this probability is zero. Moreover, for a given qubit configuration, the corresponding
control and attack circuits were always executed consecutively within the same experimental session to minimize the influence of temporal fluctuations in device calibration~\cite{Proctor2020Drift}. For processors sharing the connectivity pattern shown in Fig.~\ref{fig:IBM_top}, we employed the attack configuration introduced in Ref.~\cite{Almaguer2026}. The attack consists of applying all admissible bidirectional CNOT operations among the six working qubits surrounding the victim qubit, resulting in a total of $30$ CNOT gates. Processors exhibiting the connectivity pattern shown in Fig.~\ref{fig:miami} allow the victim qubit to be surrounded by complete square-like neighborhoods. To investigate how the size of the attacked neighborhood influences the induced bias, we considered one, two, three, and four concentric square-like neighborhoods centered at the victim qubit. For each neighborhood, all admissible bidirectional CNOT operations compatible with the processor connectivity were applied.

\begin{figure}[h!]
  \centering
  \includegraphics[width=\r\linewidth]{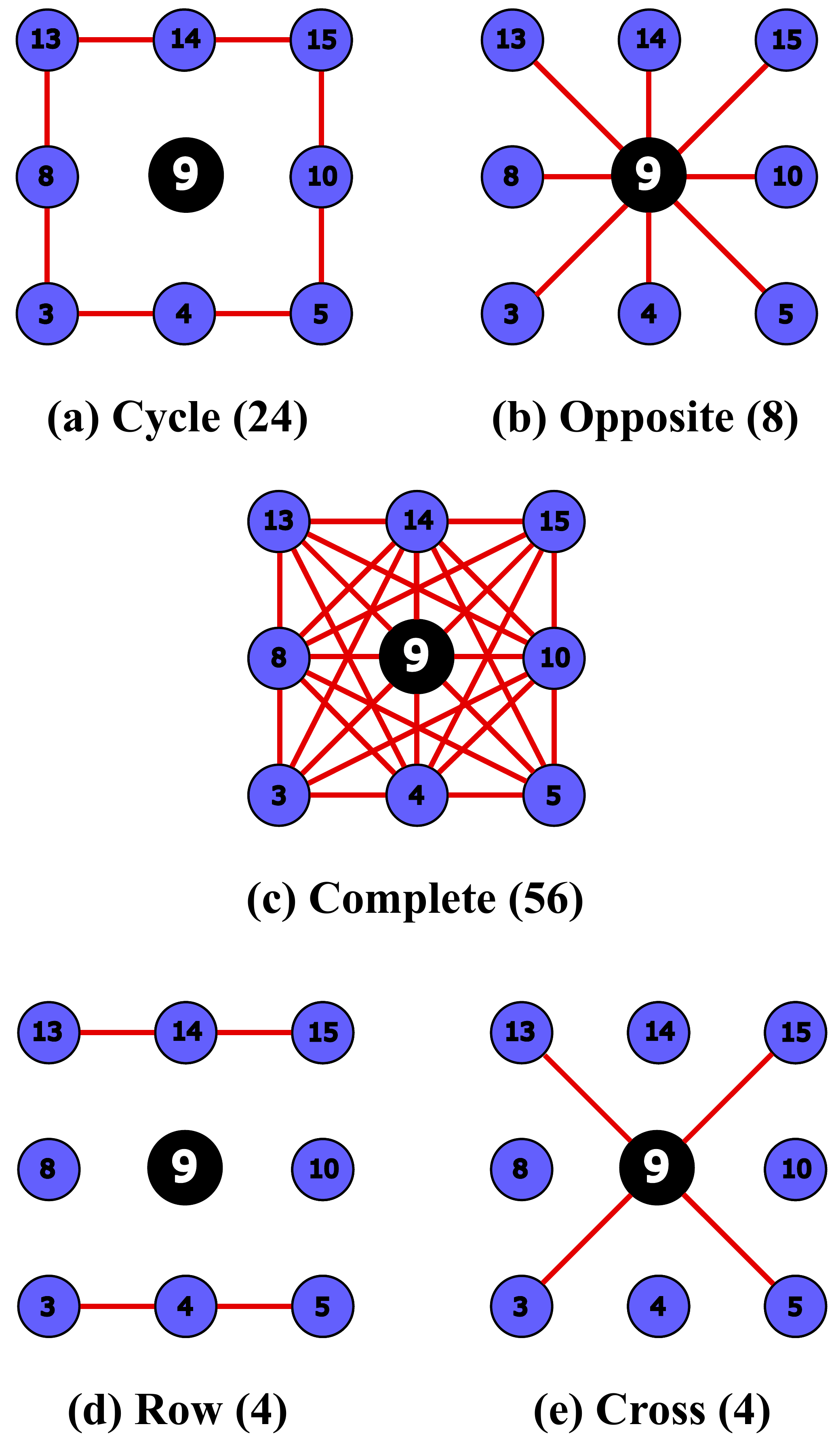}
  \caption{Interaction patterns considered for local square-like neighborhoods in the IQM Garnet processor. The neighborhood surrounding qubit $9$ is shown as a representative example; identical configurations were also implemented for victim qubits $10, 11,$ and $15$. (a) The Cycle, (b) Opposite, (c) Complete, (d) Row, and (e) Cross configurations correspond to different sets of bidirectional CNOT operations applied among neighboring working qubits. Blue and black vertices represent working qubits and the victim qubit, respectively. Red lines connect pairs of qubits between which bidirectional CNOT gates were applied. Lines crossing the victim qubit are included solely for graphical convenience and should be interpreted only as interactions between their endpoints, with the victim qubit remaining idle throughout the attack circuits. The number shown in parentheses indicates the total number of bidirectional CNOT gates employed in the corresponding configuration.
}
  \label{fig:iqm_cnots}
\end{figure}

Although the global connectivity graph of the IQM Garnet processor, shown in Fig.~\ref{fig:iqm}, differs substantially from those found in IBM devices, some regions exhibit local square-like neighborhoods similar to those available in the Miami topology. Consequently, we restricted our analysis to a single neighborhood surrounding the victim qubit and investigated several interaction patterns among the neighboring qubits. These include nearest-neighbor interactions along the neighborhood, interactions between diametrically opposite qubits, all possible pairwise interactions, and configurations involving only four selected qubits arranged either along rows or diagonals. The experiments were performed on victim qubits 9, 10, 11, and 15, which are the only qubits possessing a complete square-like neighborhood composed of eight surrounding qubits. In all cases, CNOT gates were applied in both directions between connected qubits, leaving the victim outside this interaction. The investigated interaction patterns are illustrated in Fig.~\ref{fig:iqm_cnots}.

To simplify the presentation of the experimental results, we adopt a notation inspired by Ref.~\cite{Almaguer2026} to label the different attack configurations. Labels are written in the form \texttt{cx(victim-}$n$\texttt{) C}, where \texttt{victim} identifies the attacked qubit, (C) denotes the total number of bidirectional CNOT gates employed in the corresponding interaction pattern, and (n) specifies the number of additional repetitions of the same pattern applied during the attack. When multiple interaction patterns employ the same number of bidirectional CNOT gates, an additional descriptor is appended to distinguish them (e.g., \texttt{Cross}). For example, the label \texttt{cx(9-3) 4 Cross} refers to a Cross attack implemented around victim qubit 9, consisting of four bidirectional CNOT gates and repeated three additional times.

In addition to the hardware-level characterization described above, we investigated the effects of crosstalk-induced Row Hammer attacks on a $2\rightarrow1$ and $3\rightarrow1$ quantum random access codes (QRACs)~\cite{AmbainisLeung2009QRAC,Foletto2020SequentialRAC,Anwer2020UnsharpQRAC}. Both experiments were performed on the IBM Marrakesh processor, with the victim qubit $77$ prepared in the encoded QRAC state, and physical qubits at locations 64, 65, 66, 84, 85, and 86 serving as hammer qubits.
 
Two attack positions were considered. In position A, the attack was applied before QRAC state preparation, whereas in position B, it was applied after state preparation and before the decoding measurement. This arrangement allows to investigate how the timing of the attack affects both the reconstructed quantum states and the probability of correctly retrieving the encoded classical information. Each hammer cycle consisted of bidirectional CNOT gates applied between every pair of hammer qubits, excluding the victim qubit. This all-to-all configuration therefore involved 30 CNOT gates per cycle. $N=0,5,10,$ and $20$ hammer cycles for both attack positions, using 10,000 shots per circuit were performed. For both the QRACs, all the possible codewords and recovery operations were checked. Measurements in the $X$, $Y$, and $Z$ bases were used to reconstruct the corresponding single-qubit states through quantum state tomography~\cite{James2001Measurement}, and to evaluate the average success probabilities.

\section{Results}
\label{sec:results}

\begin{figure*}[t]
\centering

\begin{minipage}{0.497\textwidth}
\centering
\includegraphics[width=\linewidth]{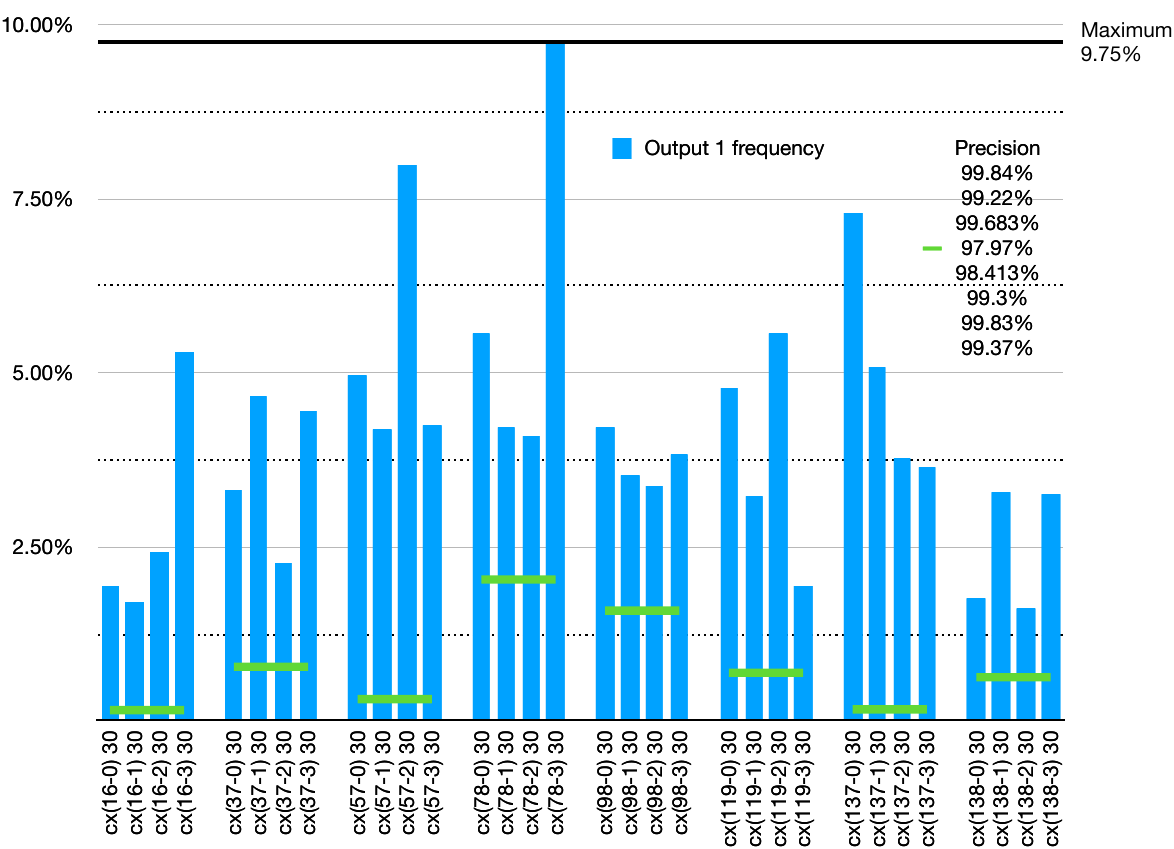}

\vspace{0.1cm}
(a) Fez
\end{minipage}
\hfill
\begin{minipage}{0.497\textwidth}
\centering
\includegraphics[width=\linewidth]{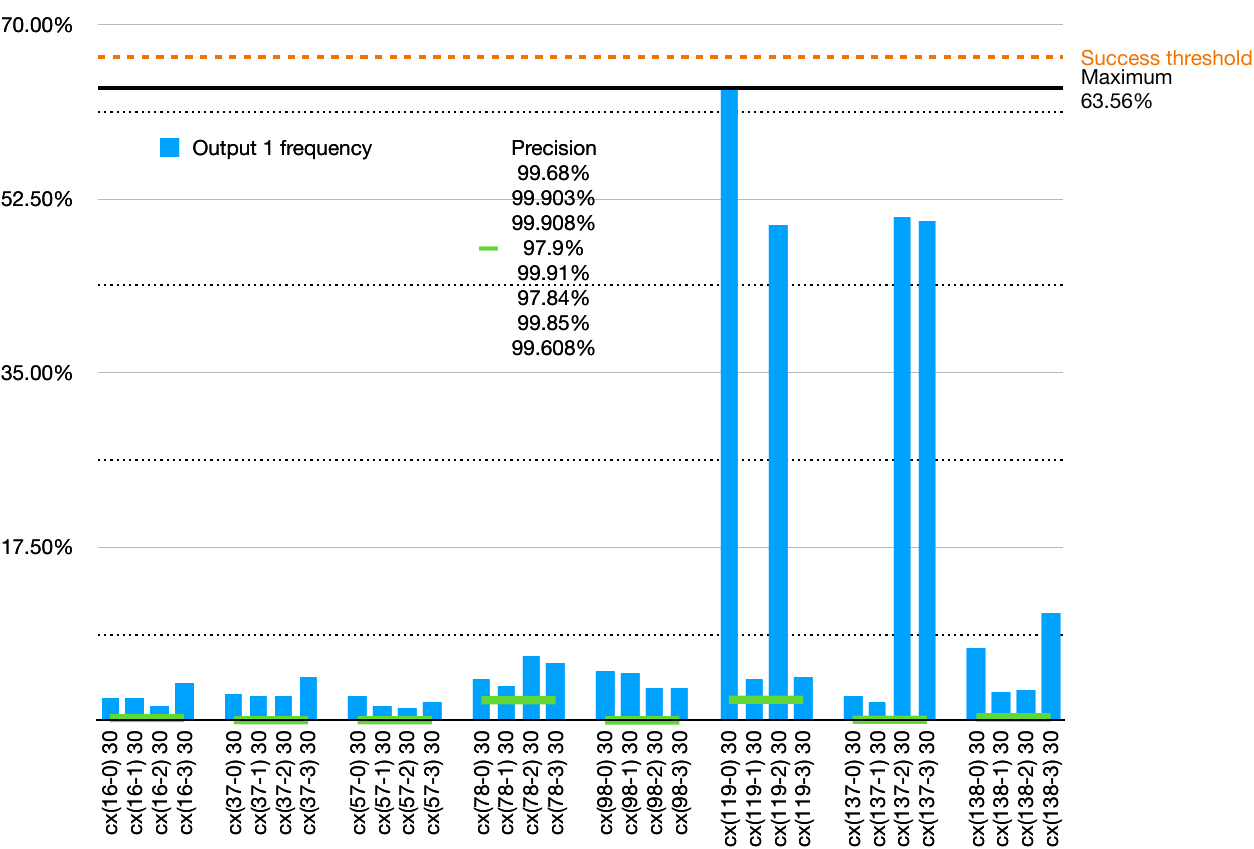}

\vspace{0.1cm}
(b) Kingston
\end{minipage}

\begin{minipage}{0.497\textwidth}
\centering
\includegraphics[width=\linewidth]{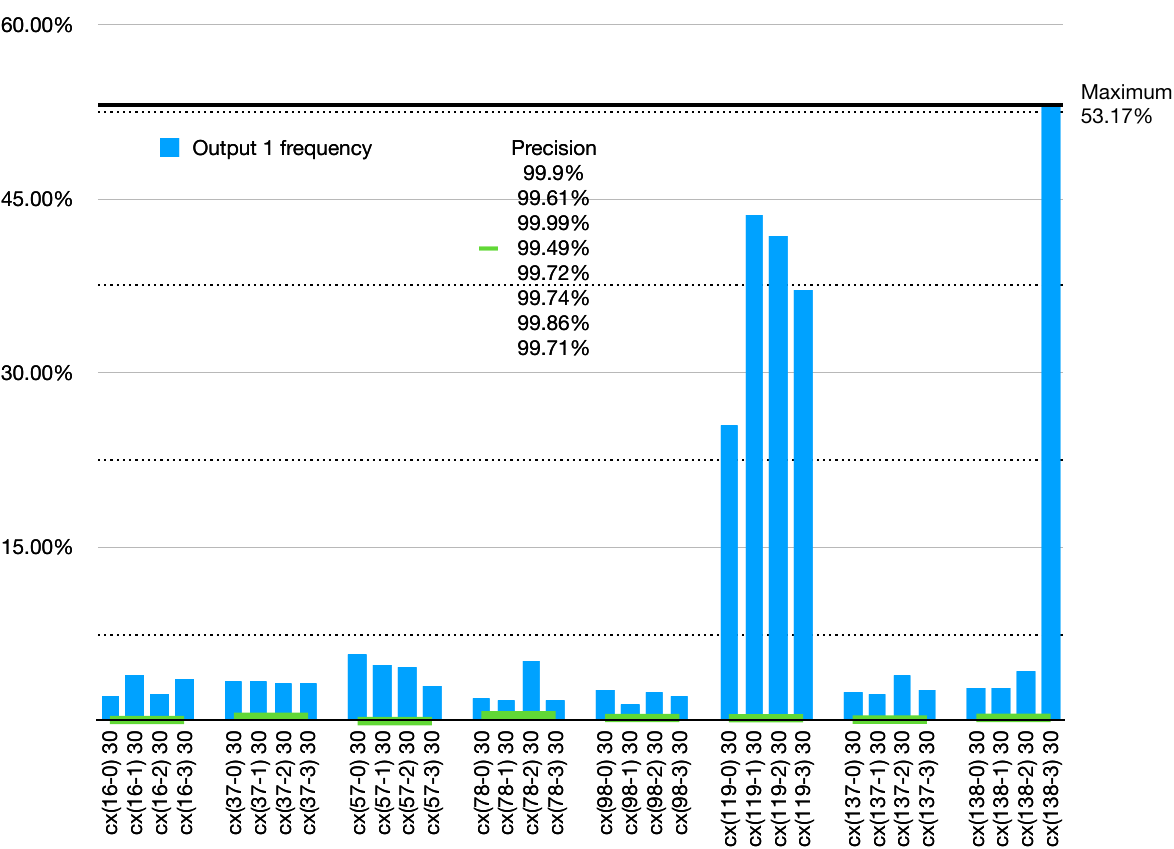}

\vspace{0.1cm}
(c) Marrakesh
\end{minipage}
\hfill
\begin{minipage}{0.497\textwidth}
\centering
\includegraphics[width=\linewidth]{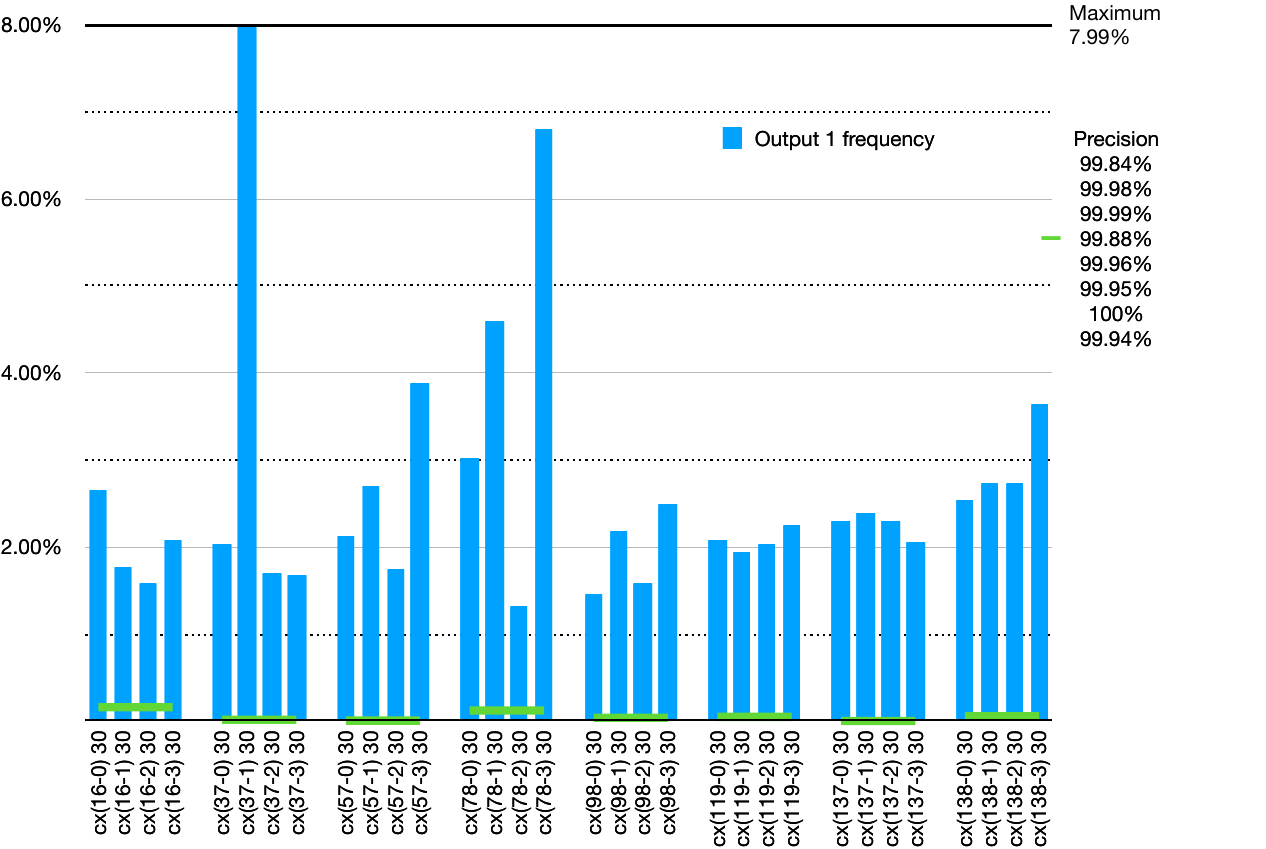}

\vspace{0.1cm}
(d) Pittsburgh
\end{minipage}

\vspace{0.3cm}

\begin{minipage}{0.497\textwidth}
\centering
\includegraphics[width=\linewidth]{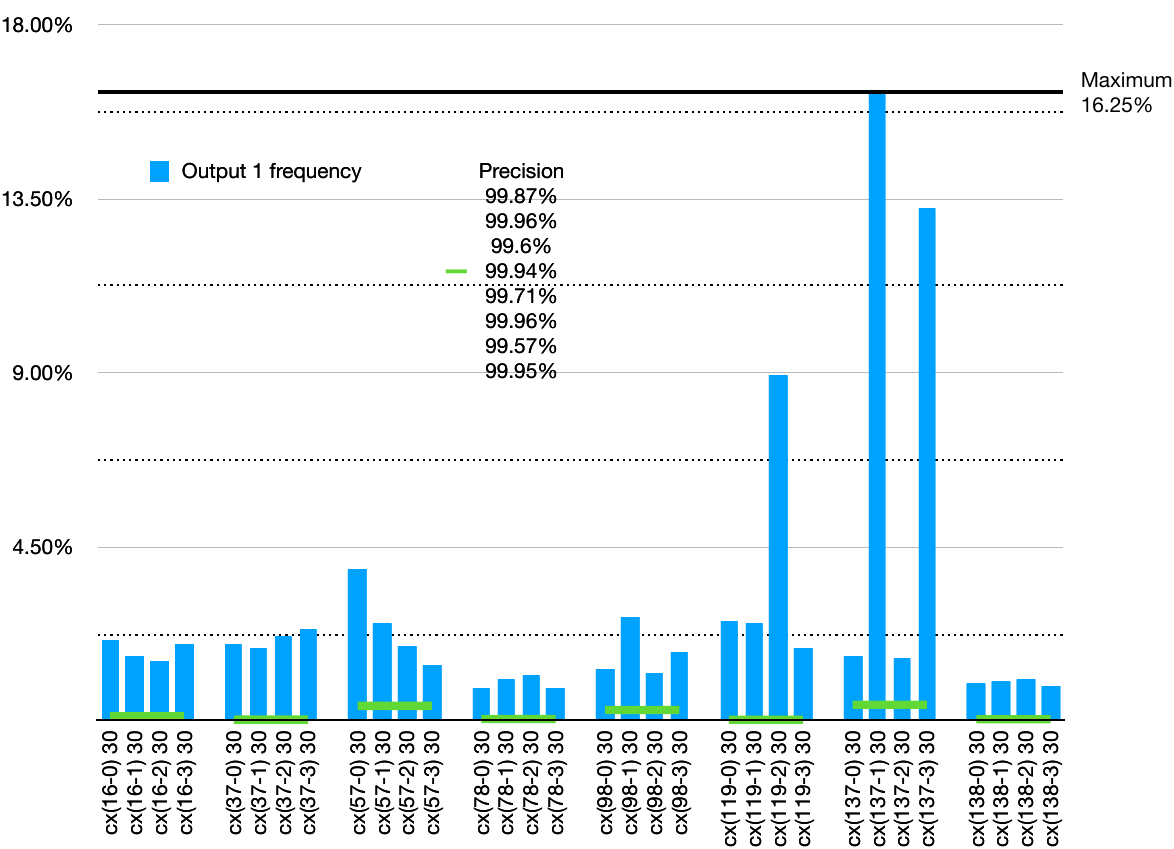}

\vspace{0.1cm}
(e) Boston
\end{minipage}

\caption{
Measured crosstalk-induced bias for IBM quantum processors sharing the connectivity graph shown in Fig.~\ref{fig:IBM_top}. Despite possessing identical coupling topologies and being subjected to the same 30-CNOT attack configuration, the devices exhibit different attack strengths and distributions. Blue bars represent the probability of measuring the victim qubit in the state $\ket{1}$ under attack, while green dashed lines indicate the control precision obtained from the corresponding control experiments.
}
\label{fig:IBMHexResults}
\end{figure*}

\begin{figure}[t]
\centering
\includegraphics[width=\r\linewidth]{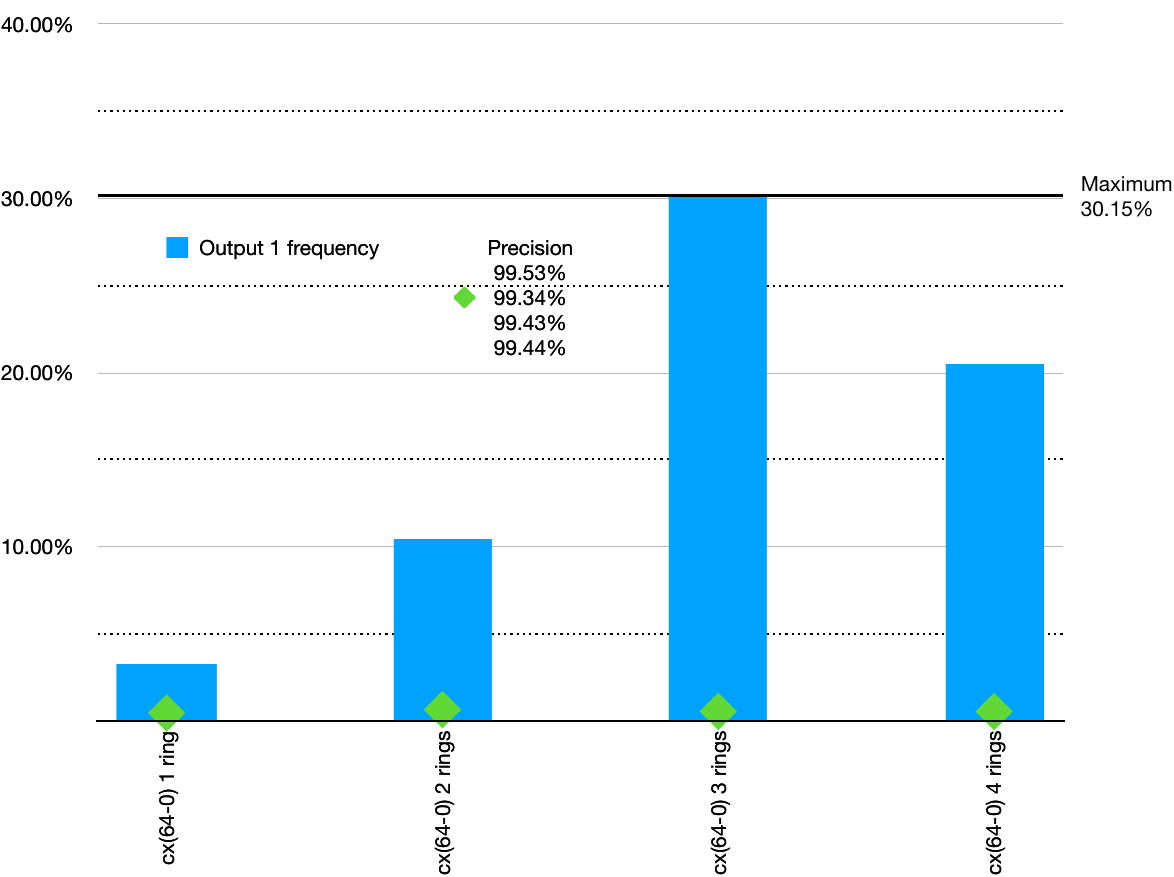}
\caption{
Measured crosstalk-induced bias for the IBM Miami processor. Configurations involving one, two, three, and four concentric square-like neighborhoods centered at victim qubit 64 were considered. Blue bars represent the probability of measuring the victim qubit in the state $\ket{1}$ under attack, while green 
points indicate the control precision obtained from the corresponding control experiments.
}
\label{fig:Miami}
\end{figure}

\begin{figure}[t]
\centering
\includegraphics[width=\r\linewidth]{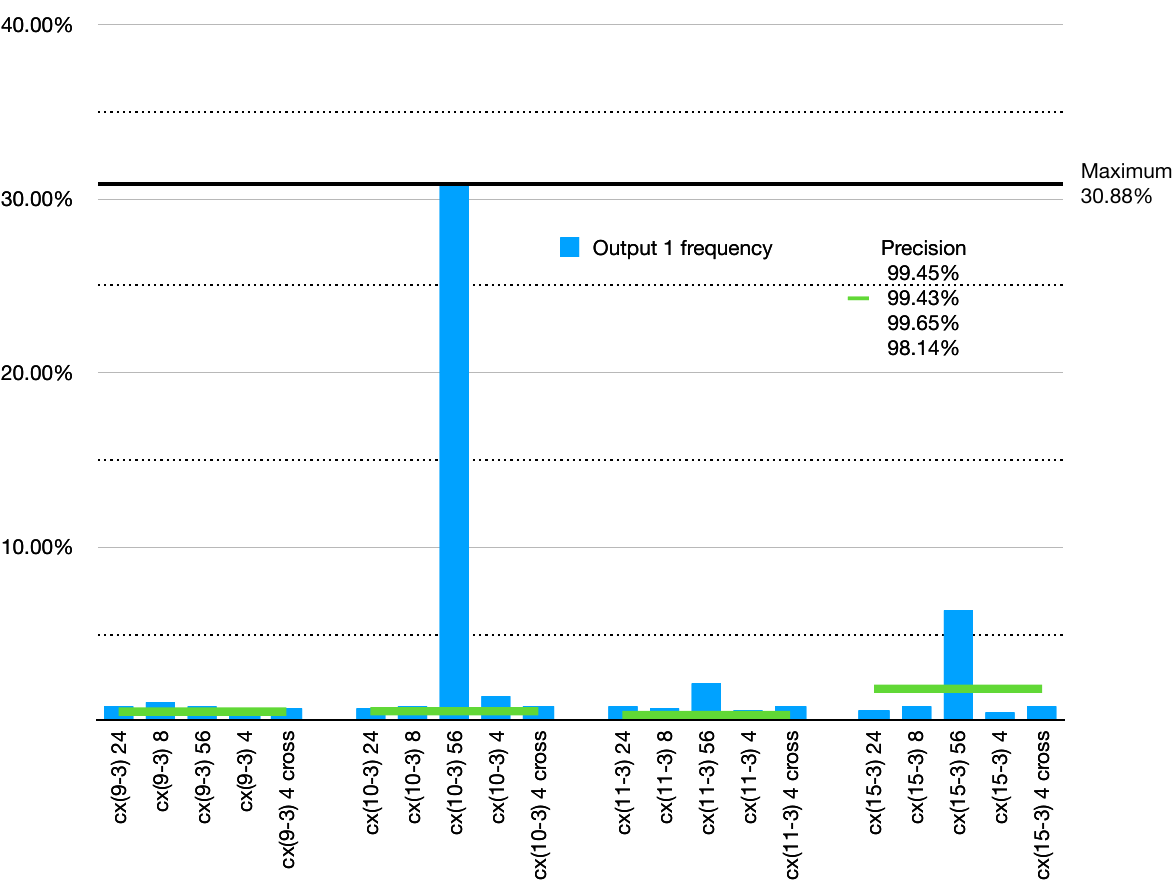}
\caption{
Measured crosstalk-induced bias for the IQM Garnet processor. Attack configurations described in Fig.~\ref{fig:iqm_cnots} were implemented around victim qubits 9, 10, 11, and 15. Blue bars represent the probability of measuring the victim qubit in the state $\ket{1}$ under attack, while green 
dashed lines  indicate the control precision obtained from the corresponding control experiments.
}
\label{fig:IQM_results}

\end{figure}
Figures~\ref{fig:IBMHexResults}--\ref{fig:IQM_results} summarize the measured crosstalk-induced bias across all investigated quantum processors and attack configurations. All devices exhibit measurable deviations in the victim qubit's measurement outcomes under at least some configurations, confirming that crosstalk-induced effects persist. However, their magnitude vary depending on the processor and reach the exploitability criterion defined in Ref.~\cite{Almaguer2026}. This suggests that attack effectiveness depends not only on device connectivity but also on hardware-specific details which is not captured in the coupling graph.

This hardware dependence is particularly evident when comparing the IBM processors in Fig.~\ref{fig:IBMHexResults}, which share the same coupling graph and are subjected to identical $30-$CNOT attack configurations. Despite these common characteristics, their responses differ with maximum measured biases ranging from below $10\%$ for Pittsburgh and Fez to $63.56\%$ for Kingston. Marrakesh also exhibits a substantial bias of up to $53.17\%$, while Boston reaches a more moderate maximum of $16.25\%$. Moreover, the crosstalk-induced bias is highly localized, with only a small subset of victim qubits and attack repetitions producing amplified deviations, while most tested configurations exhibit weaker effects. This pronounced localization indicates that processors sharing an identical connectivity graph need not have similar vulnerability profiles.

The Miami processor, whose rectangular connectivity graph allows the construction of larger square-like neighborhoods around the victim qubit exhibits a different signature. The results shown in Fig.~\ref{fig:Miami} reveal that enlarging the attacked neighborhood generally enhances the induced bias, although not monotonically. While configurations involving a single ring produce only modest deviations, the largest effect was observed for attacks employing three concentric rings, reaching $30.15\%$. Surprisingly, increasing the neighborhood to four rings resulted in a reduction of the induced bias, indicating that the attack effectiveness does not increase monotonically with the size of the attacked neighborhood.

\begin{figure}[t]\centering\includegraphics[width=\r\linewidth]{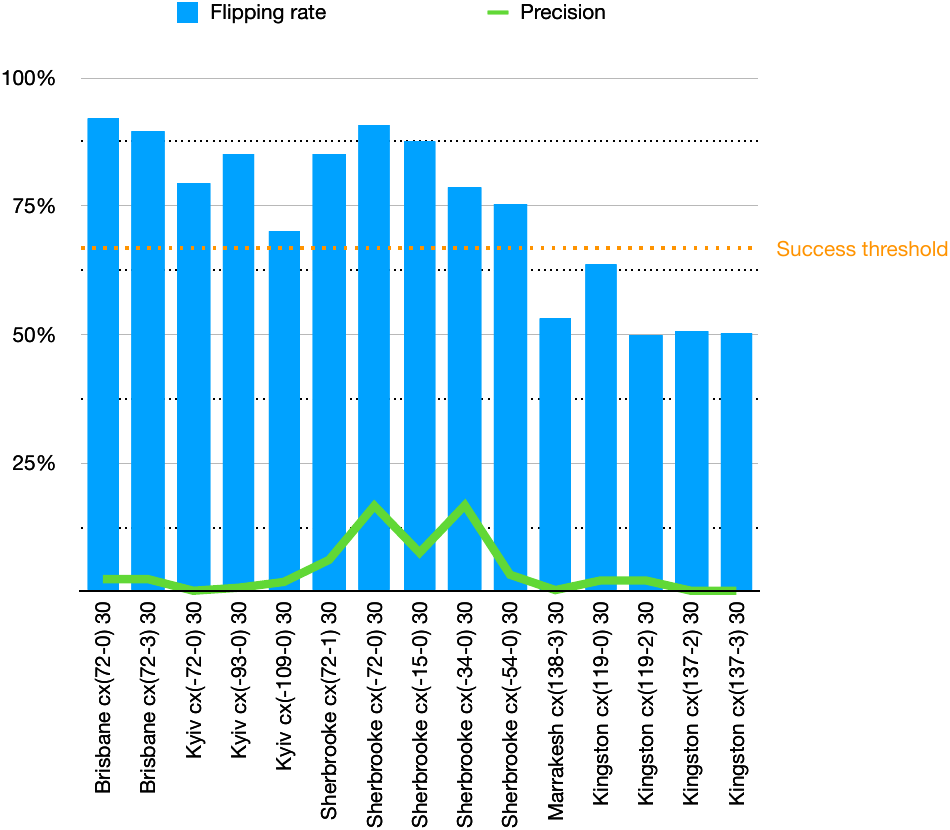}\caption{
Comparison of the strongest crosstalk-induced attacks reported for legacy IBM quantum processors in Ref.~\cite{Almaguer2026} and the strongest attack configurations identified in the present work. Blue bars represent the measured output-one probability of the victim qubit under attack, while the green markers indicate the control precision measured in the corresponding control experiments. The orange dashed line denotes the exploitability criterion introduced in Ref.~\cite{Almaguer2026}. Although most contemporary processors exhibit substantially reduced attack effectiveness, the comparison reveals that significant hardware-dependent variability persists, with the strongest configurations in the Kingston processor still approaching the previously established exploitability threshold.
}\label{fig:Merged}\end{figure}

The IQM Garnet processor exhibits a similar dependence on hardware-specific characteristics, as shown in Fig.~\ref{fig:IQM_results}. Despite all investigated victim qubits having identical local square-like neighborhoods, their responses vary based on the configuration. The Complete configuration produces the largest bias, reaching $30.88\%$ for victim qubit $10$, while inducing notably small effects for qubits $9$, $11$, and $15$. The remaining configurations: Cycle, Opposite, Row, and Cross, produce smaller deviations. These results further demonstrate that local connectivity alone does not determine the attack effectiveness, highlighting the role of hardware-specific implementation details beyond local topology. By clearly establishing that crosstalk induced attacks remain valid in the modern quantum processors, we draw comparison with the legacy IBM quantum processor from ~\cite{Almaguer2026}. It is observed that the strongest comparison can be inferred from the attack configuration shown in Fig.~\ref{fig:Merged}.

The results demonstrate that the efficiency of the attacks depend strongly on the underlying hardware and factors extending well beyond the topology itself. While several processors suppress the induced bias to values below 10\%, few of them still operate close to the exploitability regime.

\section{Impact of crosstalk-induced Row Hammer attacks on quantum random access codes}
\label{sec:qrac_attacks}

\begin{figure}[t]
    \centering
    \includegraphics[width=\linewidth]{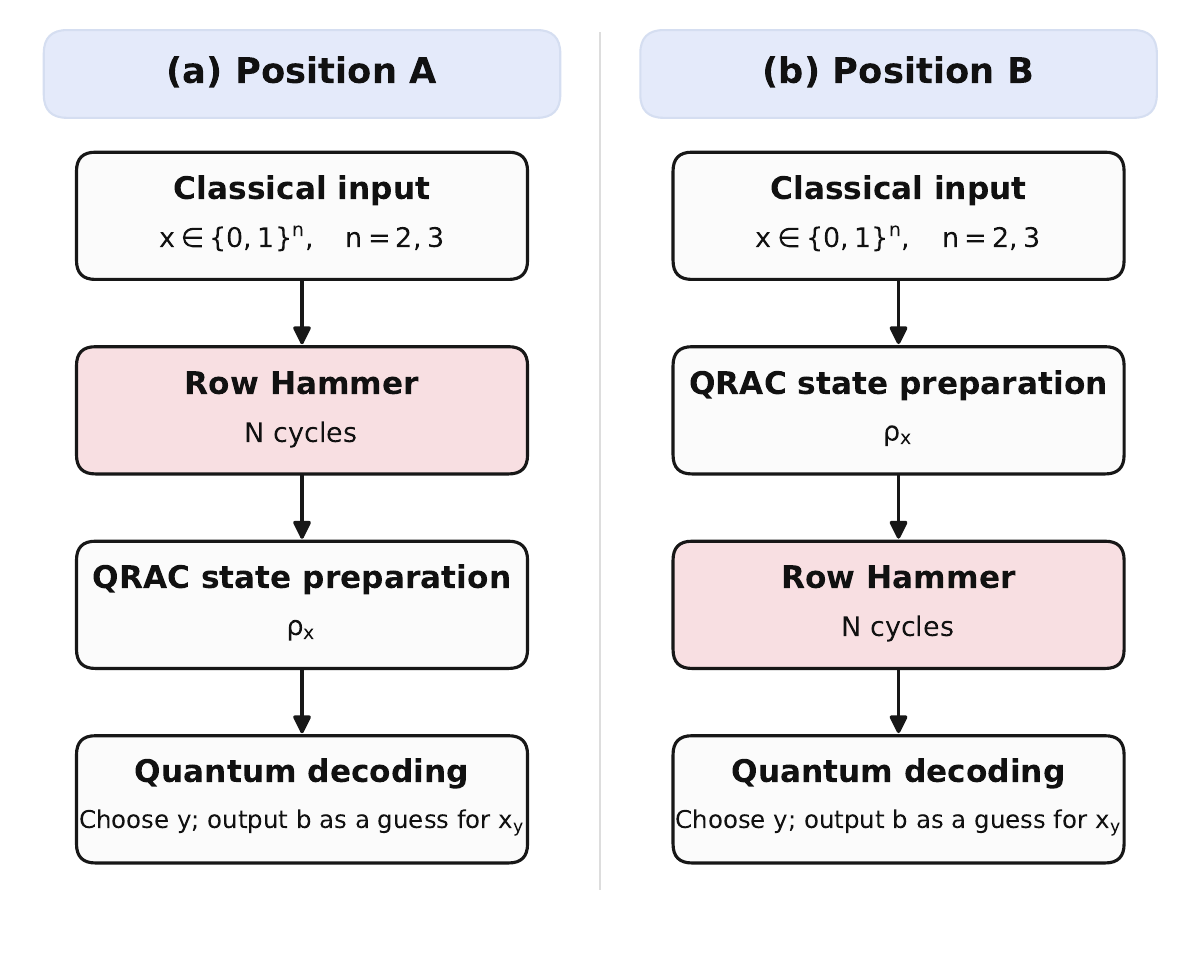}
    \caption{ 
    Schematic representation of the two row hammer attack positions considered in the QRAC experiments. In position A, the attack is applied before preparation of the QRAC state, whereas in position B, it is applied afterwards. A classical bit string $x\in\{0,1\}^n$, where $n=\{ 2,3\}$, is encoded into a qubit state $\rho_x$. The corresponding decoding quantum measurement is performed based on the bit index $y\in\{1,\ldots,n\}$ to obtain an outcome $b$ as a guess for the classical bit $x_y$.
    }
    \label{fig:row_hammer_schematic}
\end{figure}

We now investigate the impact of crosstalk-induced row hammer attacks on classical bit strings encoded as QRACs. The schematic behind the attacks applied on QRACs is shown in Fig.~\ref{fig:row_hammer_schematic}. Both QRAC protocols were implemented on the IBM Marrakesh processor using physical qubit 77 as the victim and qubits 64, 65, 66, 84, 85, and 86 as hammer qubits. Depending on the input bit value, different single qubit-rotations were applied to prepare the corresponding quantum encoded state. Each hammer cycle consisted of 30 CNOT gates applied bidirectionally between pairs of hammer qubits. We considered $N=0,5,10,$ and $20$ cycles under two attack positions. State reconstruction used measurements in the $X$, $Y$, and $Z$ bases. Each circuit was executed with 10,000 shots, yielding 96 circuits for the $2\rightarrow1$ QRAC and 96 circuits per attack position for the $3\rightarrow1$ QRAC.

\begin{figure*}[t]
    \centering
    \includegraphics[width=\textwidth]{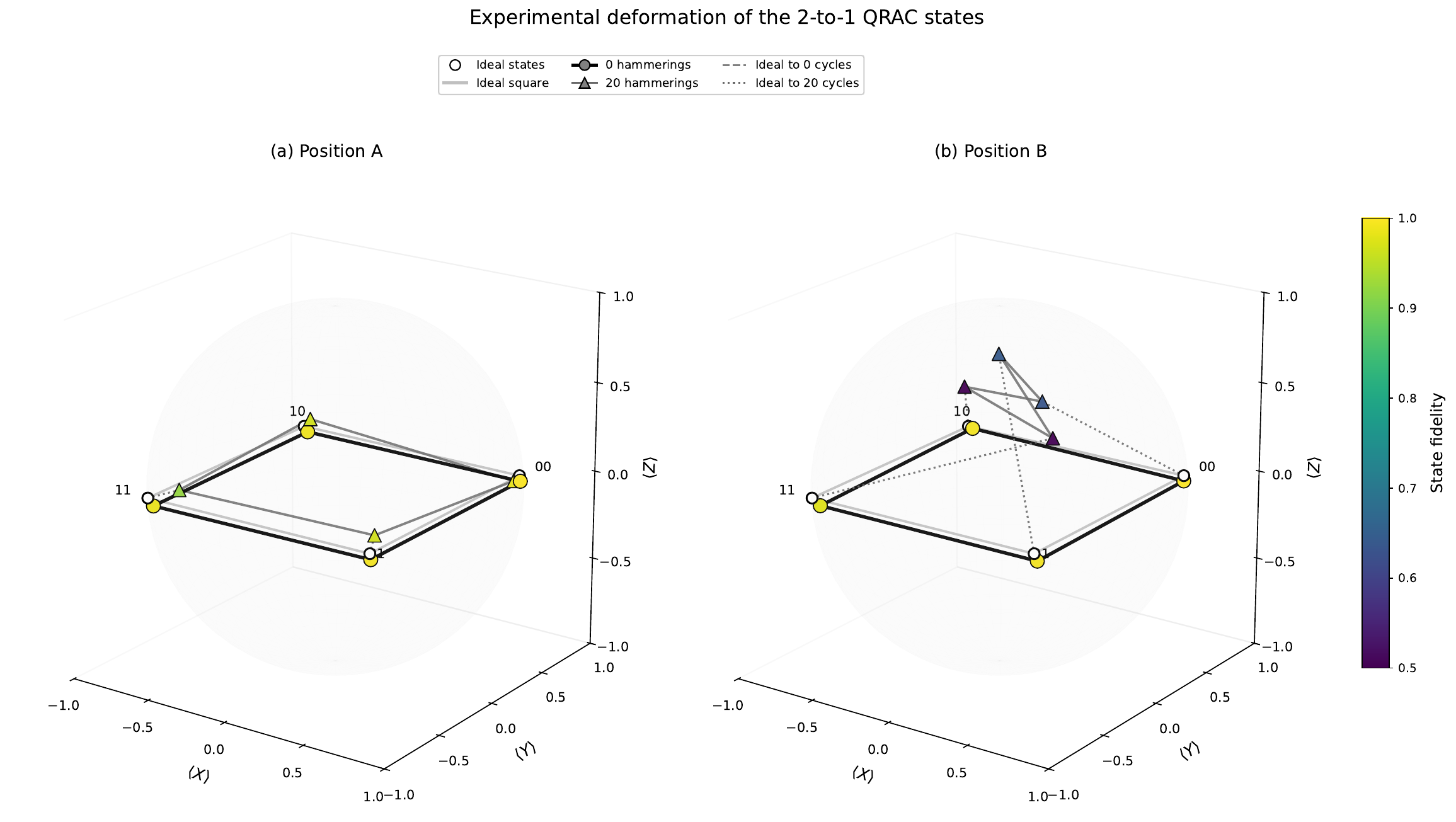}
    \caption{
    Experimental deformation of the $2\rightarrow1$ QRAC state constellation on the Bloch sphere for attack positions A (before state preparation) and B (after state preparation). White circles denote the ideal states, while colored circles and triangles represent the reconstructed states for $N=0$ and $N=20$ hammer cycles, respectively. The color of each experimental state indicates its fidelity with respect to the corresponding ideal QRAC state. Lines connecting corresponding ideal and experimental states illustrate the deformation induced by the attack.
    }
    \label{fig:cube_2_to_1}
\end{figure*}

\begin{figure*}[t]
    \centering
    \includegraphics[width=\textwidth]{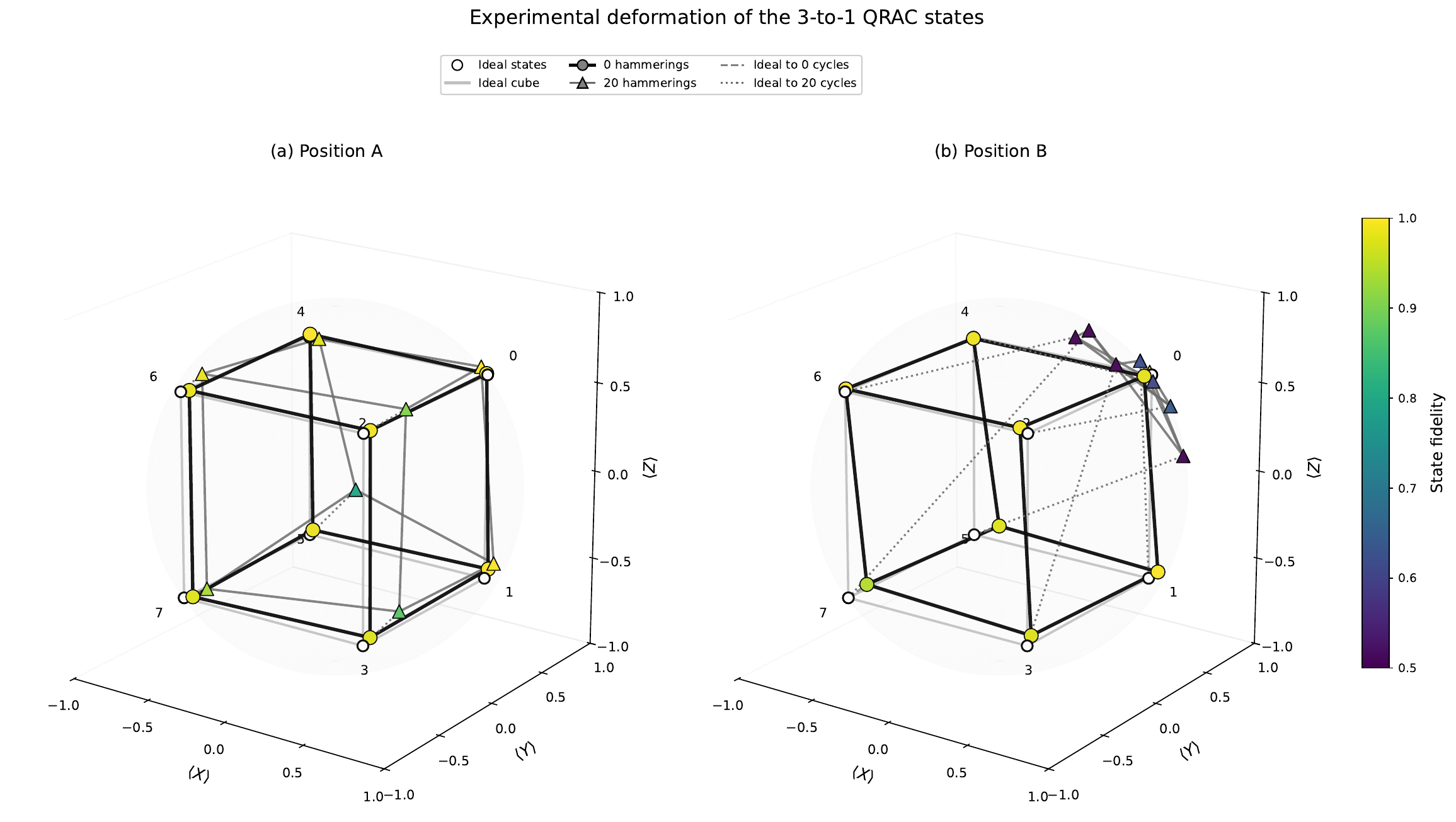}
    \caption{
    Experimental deformation of the $3\rightarrow1$ QRAC state constellation on the Bloch sphere for attack positions A (before state preparation) and B (after state preparation). White circles denote the ideal states, while colored circles and triangles represent the reconstructed states for $N=0$ and $N=20$ hammer cycles, respectively. The color of each experimental state indicates its fidelity with respect to the corresponding ideal QRAC state. Dashed lines connect corresponding ideal and experimental states to illustrate the deformation induced by the attack.
    }
    \label{fig:cube_3_to_1}
\end{figure*}

Figures~\ref{fig:cube_2_to_1} and \ref{fig:cube_3_to_1} show the experimentally reconstructed QRAC states on the Bloch sphere. The ideal states are represented by white circles, while the colored markers indicate the reconstructed states for $N=0$ and $N=20$ hammer cycles. The color scale represents the fidelity of each reconstructed state with respect to its ideal counterpart. For both protocols, the reconstructed states exhibit a more pronounced deformation for position B than for position A, particularly for $N=20$.

\begin{table}[t]
    \centering
    \begin{tabular}{lcccc}
        \toprule
        \textbf{Attack position} & $N=0$ & $N=5$ & $N=10$ & $N=20$ \\
        \midrule
        Position A & 0.8486 & 0.8188 & 0.8253 & 0.8241 \\
        Position B & 0.8484 & 0.7003 & 0.4751 & 0.5046 \\
        \bottomrule
    \end{tabular}
    \caption{
    Experimental average success probabilities of the $2\rightarrow1$ QRAC for different numbers of Row Hammer cycles, $N$, and attack positions. The ideal noiseless success probability is $P_{\mathrm{ideal}}^{2\rightarrow1}=\frac{1}{2}\left(1+\frac{1}{\sqrt{2}}\right)\approx0.8536$.
    }
    \label{tab:qrac2to1_success_probabilities}
\end{table}

\begin{table}[t]
    \centering
    \begin{tabular}{lcccc}
        \toprule
        \textbf{Attack position} & $N=0$ & $N=5$ & $N=10$ & $N=20$ \\
        \midrule
        Position A & 0.7835 & 0.7444 & 0.7325 & 0.7560 \\
        Position B & 0.7794 & 0.6852 & 0.6798 & 0.5302 \\
        \bottomrule
    \end{tabular}
    \caption{
    Experimental average success probabilities of the $3\rightarrow1$ QRAC for different numbers of Row Hammer cycles, $N$, and attack positions. The ideal noiseless success probability is $P_{\mathrm{ideal}}^{3\rightarrow1}=\frac{1}{2}\left(1+\frac{1}{\sqrt{3}}\right)\approx0.7887$.
    }
    \label{tab:qrac3to1_success_probabilities}
\end{table}

The average decoding success probabilities, reported in Tables~\ref{tab:qrac2to1_success_probabilities} and \ref{tab:qrac3to1_success_probabilities}, further illustrate the dependence of QRAC performance on the attack position. For both protocols, position A exhibits small variations in success probability across the investigated numbers of hammer cycles, with values remaining close to the respective ideal benchmarks of $0.8536$ for the $2\rightarrow1$ QRAC and $0.7887$ for the $3\rightarrow1$ QRAC. In contrast, position B leads to a stronger degradation. For the $2\rightarrow1$ QRAC, the success probability falls from $0.8484$ without hammering to $0.4751$ at $N=10$, below the $0.5$ random-guessing benchmark, before recovering slightly to $0.5046$ at $N=20$. For the $3\rightarrow1$ QRAC, it decreases from $0.7794$ without hammering to $0.5302$ at $N=20$, approaching the same benchmark. These results indicate that applying the attack after state preparation can impair the retrieval of the encoded classical information, whereas applying it before preparation has limited effect under the investigated conditions.

\section{Conclusions}
\label{sec:conclusion}

In this work, we experimentally investigated the persistence of crosstalk-induced attacks in contemporary quantum processors. By implementing specific attack configurations on six IBM superconducting processors and one IQM superconducting processor, we assessed the extent to which these vulnerabilities persist across different devices and hardware platforms. It is observed that the severity of the attack depends closely on the device specifications. The induced biases varied with processors with the largest observed value exceeding $60\%$. In particular, we note that processors with identical coupling graphs exhibit different vulnerability profiles. Hence, connectivity alone is insufficient to assess susceptibility to crosstalk-induced attacks. Other hardware-specific parameters, such as coupling strengths, qubit frequencies, residual interactions, and control-pulse characteristics, may also influence the resulting crosstalk. This highlights the need for hardware-specific characterization when assessing the effectiveness of such attacks.

The consequence of launching such attacks on QRACs during their execution on such processors is also presented. These results reveal a clear dependence of the decoding probability on the position or the time instant at which the attack is applied. Disturbances introduced before state preparation have lesser impact on decoding, whereas attacks applied after state preparation, led to a greater degradation. In particular, for one of the investigated configurations of the $2\rightarrow1$ QRAC, the average success probability dropped below the random-guessing benchmark. This demonstrates that crosstalk-induced disturbances can affect not only measurement outcomes but also the retrieval of classical information encoded in quantum states.

Overall, our study establish that crosstalk-induced attacks remain a relevant experimental concern in contemporary quantum hardware. Our results motivate security assessments that account for the physical implementation of each processor, rather than relying solely on its connectivity structure, and reinforce the need to characterize and mitigate crosstalk as quantum processors evolve.

\section{Acknowledgements}
\label{sec:ack}
VFS and ASH thank H S Karthik for valuable discussions. FAA, MP and ASH acknowledge support from project no.FENG.02.01-IP.05-0006/23, financed under the FENG Programme 2021–2027, Priority FENG.02, Measure FENG.02.01, with support from the FNP. VFS acknowledges financial support from the Coordenação de Aperfeiçoamento de Pessoal de Nível Superior (CAPES), Grant No. 23081.002220/2025-87. JM acknowledges financial support from the National Council for Scientific and Technological Development (CNPq), Grants No. 300083/2025-4, No. 409673/2022-6, and No. 421792/2022-1, and from the National Institute for the Science and Technology of Applied Quantum Computing (INCT-CQA), Grant No. 408884/2024-0. This study was financed in part by the Coordenação de Aperfeiçoamento de Pessoal de Nível Superior -- Brasil (CAPES) -- Finance Code 001.

The authors would like to express their gratitude to IBM and IQM, which proved valuable in supporting the experimental aspects of this work. The authors also wish to clarify these computing systems were employed solely as a technical platform to facilitate the scientific investigation presented. Their usage does not imply any criticism or negative assessment of the company's products or services. This research is driven purely by academic curiosity and aims to contribute to the broader understanding of system security in computing environments.

\bibliographystyle{apsrev4-2}
\bibliography{References}

@article{Kim2023,
  author  = {Kim, Youngseok and Eddins, Andrew and
             Anand, Sajant and Wei, Ken Xuan and
             van den Berg, Ewout and Rosenblatt, Sami and
             Nayfeh, Hasan and Wu, Yantao and
             Zaletel, Michael and Temme, Kristan and
             Kandala, Abhinav},
  title   = {Evidence for the utility of quantum computing before fault tolerance},
  journal = {Nature},
  volume  = {618},
  pages   = {500--505},
  year    = {2023},
  doi     = {10.1038/s41586-023-06096-3}
}

@article{Zhao2022,
  author  = {Zhao, Peng and Linghu, Kehuan and Li, Zhiyuan and
             Xu, Peng and Wang, Ruixia and Xue, Guangming and
             Jin, Yirong and Yu, Haifeng},
  title   = {Quantum Crosstalk Analysis for Simultaneous Gate Operations on Superconducting Qubits},
  journal = {PRX Quantum},
  volume  = {3},
  pages   = {020301},
  year    = {2022},
  doi     = {10.1103/PRXQuantum.3.020301}
}

@article{Almaguer2026,
  title   = {Configuration dependent crosstalk amplification in superconducting quantum processors},
  author  = {Almaguer-Angeles, Fernando and Ruas-Dieguez, Pedro and Shenoy Hejmadi, Akshata and Paw{\l}owski, Marcin},
  journal = {Scientific Reports},
  year    = {2026},
  doi     = {10.1038/s41598-026-70466-w},
  url     = {https://doi.org/10.1038/s41598-026-70466-w}
}

@article{Wang2019EARAC,
  author  = {Wang, Xiao-Run and Wu, Lu-Yan and Liu, Chen-Xi and
             Liu, Tong-Jun and Li, Jian and Wang, Qin},
  title   = {Experimental generation of entanglement-assisted
             quantum random access code},
  journal = {Physical Review A},
  volume  = {99},
  pages   = {052313},
  year    = {2019},
  doi     = {10.1103/PhysRevA.99.052313}
}

@article{Harper2025,
  author  = {Harper, Benjamin and Tonekaboni, Behnam and
             Goldozian, Bahar and Sevior, Martin and Usman, Muhammad},
  title   = {Crosstalk Attacks and Defence in a Shared Quantum Computing Environment},
  journal = {Advanced Quantum Technologies},
  volume  = {8},
  number  = {10},
  pages   = {e2500009},
  year    = {2025},
  doi     = {10.1002/qute.202500009}
}

@article{Gross2024,
  author  = {Gross, Jonathan A. and Genois, {\'E}lie and
             Debroy, Dripto M. and Zhang, Yaxing and
             Mruczkiewicz, Wojciech and Cian, Ze-Pei and
             Jiang, Zhang},
  title   = {Characterizing coherent errors using matrix-element amplification},
  journal = {npj Quantum Information},
  volume  = {10},
  year    = {2024},
  doi     = {10.1038/s41534-024-00917-7}
}

@article{Sarovar2020,
  author  = {Sarovar, Mohan and Proctor, Timothy and
             Rudinger, Kenneth and Young, Kevin and
             Nielsen, Erik and Blume-Kohout, Robin},
  title   = {Detecting crosstalk errors in quantum information processors},
  journal = {Quantum},
  volume  = {4},
  pages   = {321},
  year    = {2020},
  doi     = {10.22331/q-2020-09-11-321}
}

@article{KettererWellens2023,
  author  = {Ketterer, Andreas and Wellens, Thomas},
  title   = {Characterizing Crosstalk of Superconducting Transmon Processors},
  journal = {Physical Review Applied},
  volume  = {20},
  number  = {3},
  pages   = {034065},
  year    = {2023},
  doi     = {10.1103/PhysRevApplied.20.034065}
}

@article{AshSaki2020Crosstalk,
  author  = {Ash-Saki, Abdullah and Alam, Mahabubul and Ghosh, Swaroop},
  title   = {Experimental Characterization, Modeling, and Analysis
             of Crosstalk in a Quantum Computer},
  journal = {IEEE Transactions on Quantum Engineering},
  volume  = {1},
  pages   = {3101406},
  year    = {2020},
  doi     = {10.1109/TQE.2020.3023338}
}

@article{Gambetta2012Addressability,
  author  = {Gambetta, Jay M. and C{\'o}rcoles, A. D. and Merkel,
             Seth T. and Johnson, Blake R. and Smolin, John A.
             and Chow, Jerry M. and Ryan, Colm A. and Rigetti,
             Chad and Poletto, S. and Ohki, Thomas A. and
             Ketchen, Mark B. and Steffen, Matthias},
  title   = {Characterization of Addressability by Simultaneous
             Randomized Benchmarking},
  journal = {Physical Review Letters},
  volume  = {109},
  pages   = {240504},
  year    = {2012},
  doi     = {10.1103/PhysRevLett.109.240504}
}

@article{Rudinger2021,
  author  = {Rudinger, Kenneth and Hogle, Craig W. and
             Naik, Ravi K. and Hashim, Akel and
             Lobser, Daniel and Santiago, David I. and
             Grace, Matthew D. and Nielsen, Erik and
             Proctor, Timothy and Seritan, Stefan and
             Clark, Susan M. and Blume-Kohout, Robin and
             Siddiqi, Irfan and Young, Kevin C.},
  title   = {Experimental Characterization of Crosstalk Errors
             with Simultaneous Gate Set Tomography},
  journal = {PRX Quantum},
  volume  = {2},
  pages   = {040338},
  year    = {2021},
  doi     = {10.1103/PRXQuantum.2.040338}
}

@article{James2001Measurement,
  author  = {James, Daniel F. V. and Kwiat, Paul G. and
             Munro, William J. and White, Andrew G.},
  title   = {Measurement of qubits},
  journal = {Physical Review A},
  volume  = {64},
  number  = {5},
  pages   = {052312},
  year    = {2001},
  doi     = {10.1103/PhysRevA.64.052312}
}

@article{Kosen2024,
  author  = {Kosen, Sandoko and Li, Hang-Xi and Rommel, Marcus and
             Rehammar, Robert and Caputo, Marco and Gronberg, Leif and
             Fernandez-Pendas, Jorge and Kockum, Anton Frisk and
             Biznarova, Janka and Chen, Liangyu and Križan, Christian and
             Nylander, Andreas and Osman, Amr and Roudsari, Anita Fadavi and
             Shiri, Daryoush and Tancredi, Giovanna and Govenius, Joonas and
             Bylander, Jonas},
  title   = {Signal Crosstalk in a Flip-Chip Quantum Processor},
  journal = {PRX Quantum},
  volume  = {5},
  pages   = {030350},
  year    = {2024},
  doi     = {10.1103/PRXQuantum.5.030350}
}

@article{Chamberland2020HeavyHex,
  author  = {Chamberland, Christopher and Zhu, Guanyu and
             Yoder, Theodore J. and Hertzberg, Jared B. and
             Cross, Andrew W.},
  title   = {Topological and Subsystem Codes on Low-Degree
             Graphs with Flag Qubits},
  journal = {Physical Review X},
  volume  = {10},
  pages   = {011022},
  year    = {2020},
  doi     = {10.1103/PhysRevX.10.011022}
}

@article{HetenyiWootton2024,
  author  = {Het{\'e}nyi, Bence and Wootton, James R.},
  title   = {Creating Entangled Logical Qubits in the Heavy-Hex
             Lattice with Topological Codes},
  journal = {PRX Quantum},
  volume  = {5},
  pages   = {040334},
  year    = {2024},
  doi     = {10.1103/PRXQuantum.5.040334}
}

@article{Foletto2020SequentialRAC,
  author  = {Foletto, Giulio and Calderaro, Luca and
             Vallone, Giuseppe and Villoresi, Paolo},
  title   = {Experimental demonstration of sequential
             quantum random access codes},
  journal = {Physical Review Research},
  volume  = {2},
  number  = {3},
  pages   = {033205},
  year    = {2020},
  doi     = {10.1103/PhysRevResearch.2.033205}
}

@article{Anwer2020UnsharpQRAC,
  author  = {Anwer, Hammad and Muhammad, Sadiq and
             Cherifi, Walid and Miklin, Nikolai and
             Tavakoli, Armin and Bourennane, Mohamed},
  title   = {Experimental Characterization of Unsharp Qubit
             Observables and Sequential Measurement
             Incompatibility via Quantum Random Access Codes},
  journal = {Physical Review Letters},
  volume  = {125},
  number  = {8},
  pages   = {080403},
  year    = {2020},
  doi     = {10.1103/PhysRevLett.125.080403}
}

@article{Proctor2020Drift,
  author  = {Proctor, Timothy and Revelle, Melissa and
             Nielsen, Erik and Rudinger, Kenneth and
             Lobser, Daniel and Maunz, Peter and
             Blume-Kohout, Robin and Young, Kevin},
  title   = {Detecting and tracking drift in quantum
             information processors},
  journal = {Nature Communications},
  volume  = {11},
  pages   = {5396},
  year    = {2020},
  doi     = {10.1038/s41467-020-19074-4}
}

@article{Nation2021Scalable,
  author  = {Nation, Paul D. and Kang, H. and
             Sundaresan, N. and Gambetta, Jay M.},
  title   = {Scalable mitigation of measurement errors
             on quantum computers},
  journal = {PRX Quantum},
  volume  = {2},
  pages   = {040326},
  year    = {2021},
  doi     = {10.1103/PRXQuantum.2.040326}
}

@inproceedings{Seif2024,
  author    = {Seif, Alireza and Liao, Haoran and Tripathi, Vinay and
               Krsulich, Kevin and Malekakhlagh, Moein and Amico, Mirko and
               Jurcevic, Petar and Javadi-Abhari, Ali},
  title     = {Suppressing Correlated Noise in Quantum Computers via Context-Aware Compiling},
  booktitle = {Proceedings of the 2024 ACM/IEEE 51st Annual International Symposium on Computer Architecture (ISCA)},
  pages     = {310--324},
  year      = {2024},
  doi       = {10.1109/ISCA59077.2024.00031}
}

@misc{heng2024estimating,
  title         = {Estimating the Effect of Crosstalk Error on Circuit Fidelity Using Noisy Intermediate-Scale Quantum Devices},
  author        = {Heng, Sovanmonynuth and Go, Myeongseong and Han, Youngsun},
  year          = {2024},
  eprint        = {2402.06952},
  archivePrefix = {arXiv},
  primaryClass  = {quant-ph},
  url           = {https://arxiv.org/abs/2402.06952}
}

@article{perrin2024mitigating,
  title     = {Mitigating crosstalk errors by randomized compiling: Simulation of the BCS model on a superconducting quantum computer},
  author    = {Perrin, Hugo and Scoquart, Thibault and Shnirman, Alexander and Schmalian, J{\"o}rg and Snizhko, Kyrylo},
  journal   = {Physical Review Research},
  volume    = {6},
  number    = {1},
  pages     = {013142},
  year      = {2024},
  doi       = {10.1103/PhysRevResearch.6.013142}
}

@article{Preskill2018,
  doi = {10.22331/q-2018-08-06-79},
  url = {https://doi.org/10.22331/q-2018-08-06-79},
  title = {Quantum {C}omputing in the {NISQ} era and beyond},
  author = {Preskill, John},
  journal = {{Quantum}},
  issn = {2521-327X},
  publisher = {{Verein zur F{\"{o}}rderung des Open Access Publizierens in den Quantenwissenschaften}},
  volume = {2},
  pages = {79},
  month = aug,
  year = {2018}
}

@article{Tavakoli2015,
  author  = {Tavakoli, Armin and Marques, Breno and
             Pawlowski, Marcin and Bourennane, Mohamed},
  title   = {Quantum Random Access Codes Using Single {D}-Level Systems},
  journal = {Physical Review Letters},
  volume  = {114},
  pages   = {170502},
  year    = {2015},
  doi     = {10.1103/PhysRevLett.114.170502}
}

@misc{AmbainisLeung2009QRAC,
  author       = {Ambainis, Andris and Leung, Debbie and
                  Mancinska, Laura and Ozols, Maris},
  title        = {Quantum Random Access Codes with Shared Randomness},
  year         = {2009},
  eprint       = {0810.2937},
  archivePrefix = {arXiv},
  primaryClass = {quant-ph},
  url          = {https://arxiv.org/abs/0810.2937}
}

@article{MehraKalev2026DefendingCrosstalk,
  title   = {Towards defending crosstalk-mediated attacks in multi-tenant quantum computing},
  author  = {Mehra, Devika and Kalev, Amir},
  journal = {Physica Scripta},
  volume  = {101},
  number  = {9},
  pages   = {095102},
  year    = {2026},
  doi     = {10.1088/1402-4896/ae4429},
  url     = {https://doi.org/10.1088/1402-4896/ae4429}
}

@inproceedings{ShubhaFarheen2025PulseLevelCrosstalk,
  title     = {Pulse-Level Simulation of Crosstalk Attacks on Superconducting Quantum Hardware},
  author    = {Shubha, Syed Emad Uddin and Farheen, Tasnuva},
  booktitle = {2025 IEEE International Conference on Quantum Computing and Engineering (QCE)},
  year      = {2025},
  pages = {91--94},
  doi = {10.1109/QCE65121.2025.10300},
  url       = {https://doi.org/10.1109/QCE65121.2025.10300}
}

@inproceedings{Arellano2025QubitVise,
  title     = {QubitVise: Double-Sided Crosstalk Attack in Superconducting Quantum Computers},
  author    = {Aranguren Arellano, Adriana and Xie, He and Szefer, Jakub},
  booktitle = {2025 IEEE International Conference on Quantum Computing and Engineering (QCE)},
  year      = {2025},
  pages     = {50--53},
  doi       = {10.1109/QCE65121.2025.10292},
  url       = {https://doi.org/10.1109/QCE65121.2025.10292}
}

@article{Kang2025TimeAdaptiveCrosstalk,
  title   = {Time-adaptive single-shot crosstalk detector on a superconducting quantum computer},
  author  = {Kang, Haiyue and Harper, Benjamin and Usman, Muhammad and Sevior, Martin},
  journal = {Physical Review Applied},
  volume  = {24},
  pages   = {024034},
  year    = {2025},
  doi     = {10.1103/mbkk-j663},
  url     = {https://doi.org/10.1103/mbkk-j663}
}

@article{ChenWang2026XYCrosstalk,
  title   = {Scalable suppression of {XY} crosstalk by pulse-level control in superconducting quantum processors},
  author  = {Chen, Hui-Hang and Wang, Chiao-Hsuan},
  journal = {Physical Review Applied},
  volume  = {26},
  pages   = {034001},
  year    = {2026},
  doi     = {10.1103/v8p3-qh5k},
  url     = {https://doi.org/10.1103/v8p3-qh5k}
}

@article{Wesdorp2026SimultaneousSingleQubitCrosstalk,
  title   = {Mitigating crosstalk errors for simultaneous single-qubit gates on a superconducting quantum processor},
  author  = {Wesdorp, Jaap J. and others},
  journal = {PRX Quantum},
  year    = {2026},
  note    = {Accepted for publication on 23 September 2026},
  doi     = {10.1103/qhss-mgkt}
}

@article{AdhikariHalunen2025HardwareVulnerabilities,
  title   = {Review of vulnerabilities in quantum hardware},
  author  = {Adhikari, Subarna and Halunen, Kimmo},
  journal = {Journal of Physics Communications},
  volume  = {9},
  number  = {11},
  pages   = {112001},
  year    = {2025},
  doi     = {10.1088/2399-6528/ae1652},
  url     = {https://doi.org/10.1088/2399-6528/ae1652}
}

@misc{jakub24,
    author = {Yizhuo Tan, Navnil Choudhury, Kanad Basu and Jakub Szefer},
    title = {QubitHammer Attacks: Qubit Flipping Attacks in Multi-tenant Superconducting Quantum Computers},
    note = {arXiv.2504.07875},
    url = {https://arxiv.org/abs/2504.07875}
}

@article{karthik25,
  title = {Noise-resilient quantum random access codes},
  author = {Karthik, H. S. and G{\'o}mez, S. and Quinteros, F. M. and H., Akshata Shenoy and Paw{\l{}}owski, M. and Walborn, S. P. and Lima, G. and G{\'o}mez, E. S.},
  journal = {Phys. Rev. A},
  volume = {111},
  issue = {3},
  pages = {032613},
  numpages = {8},
  year = {2025},
  month = {Mar},
  publisher = {American Physical Society},
  doi = {10.1103/PhysRevA.111.032613},
  url = {https://link.aps.org/doi/10.1103/PhysRevA.111.032613}
}
\end{document}